\documentclass[11pt]{article}
\usepackage{amssymb, amsmath, amsthm, epsf, epsfig}

\newlength{\originalbase}
\newcommand{\ep}{\varepsilon}
\newcommand{\E}{{\rm E}}

\newcommand{\Vbar}{\overline V}
\newcommand{\Xbar}{\overline X}
\newcommand{\Ybar}{\overline Y}

\newcommand{\cE}{{\cal E}}

\newcommand{\cX}{{\cal X}}
\newcommand{\cY}{{\cal Y}}
\newcommand{\cZ}{{\cal Z}}
\newcommand{\diag}{{\rm diag}}

\makeatletter
\def\multilimits@{\bgroup
  \Let@
  \restore@math@cr
  \default@tag
 \baselineskip\fontdimen10 \scriptfont\tw@
 \advance\baselineskip\fontdimen12 \scriptfont\tw@
 \lineskip\thr@@\fontdimen8 \scriptfont\thr@@
 \lineskiplimit\lineskip
 \vbox\bgroup\ialign\bgroup\hfil$\m@th\scriptstyle{##}$\hfil\crcr}
\def\Sb{_\multilimits@}
\def\Sp{^\multilimits@}
\def\endSb{\crcr\egroup\egroup\egroup}

\makeatother

\begin{document}
\newtheorem{theorem}{Theorem}[section]
\newtheorem{prop}[theorem]{Proposition}
\newtheorem{lemma}[theorem]{Lemma}
\newtheorem{corollary}[theorem]{Corollary}
\newtheorem{guess}[theorem]{Conjecture}
\newtheorem{conjecture}[theorem]{Conjecture}
\begin{center}
{\Large\bf Quantum Data Compression of Ensembles of Mixed States with 
Commuting Density Operators}\\ \vspace{0.3cm}
{Gerhard Kramer, Bell Labs 2C-174, Murray Hill NJ 07974}\\
{Serap A. Savari, Bell Labs 2C-451, Murray Hill NJ 07974}\\
\end{center}

\begin{abstract}
We provide a rate distortion interpretation of the problem of quantum data
compression of ensembles of mixed states with commuting density operators.
There are two versions of this problem.
In the {\em visible} case the sequence of states is available to the
encoder and in the {\em blind} or {\em hidden} 
case the encoder may access only a sequence of measurements.
We find the exact optimal compression rates for both the visible and
hidden cases. Our analysis includes the scenario in which
asymptotic reconstruction is imperfect.
\end{abstract}

\section{Introduction}
Claude Shannon created the foundations of information theory,
a mathematical theory of communication,
in his landmark 1948 paper \cite{shannon}.
However, until fairly recently few attempts were made to study the 
transmission and processing of quantum states.
The excellent survey paper \cite{bennett} provides considerable motivation
for the study of quantum information theory.
Important application areas include quantum cryptographic protocols that are 
more secure than and quantum computers that are dramatically faster than their
classical counterparts.

The first problem that Shannon addressed in \cite{shannon} was the ultimate
data compression achievable on the output of a discrete information source.
Shannon initially considered the set of encoding rules for which the source
sequence can be perfectly retrieved from the encoded sequence, at least
with high probability.
For any discrete, stationary, and ergodic source, Shannon defined the
{\em entropy} of the source as a function of the probabilities of the source
and demonstrated that the minimum achievable average number of code symbols
per source symbol is the entropy of the source.
Later in another paper \cite{shannon59}, Shannon also treated the problem of
encoding a source given a {\em fidelity criterion} or a 
{\em measure of the distortion} for a representation of the source output.
The goal in this case is to minimize the expected distortion attainable at 
a particular rate.
For a wide class of distortion measures and source models, Shannon provided
a generalization of the source entropy, known as the {\em rate distortion
function}, which establishes the exact trade off between the distortion level
and the compression rate.  

An important problem in the field of quantum information theory is the
generalization of classical results on data compression to the quantum domain.
To our knowledge, the literature thus far treats quantum analogs of discrete,
memoryless sources and assumes that the reconstruction must have 
arbitrarily high fidelity in the limit as the source string length approaches
infinity.

In order to describe a discrete, memoryless quantum source, we must first
define {\em pure} and {\em mixed} quantum states.
The state space of a quantum system is a complete description of the properties
of the particles in the system.
It includes information about positions, momentums, polarizations, spins, and
so on.
The state space is commonly modelled by a Hilbert space of wave functions.
The mathematical tools used for the study of quantum information systems are
finite dimensional complex vector spaces with an inner product that are
spanned by abstract wave functions.
A thorough discussion of mathematical conventions and terminology which are
standard in quantum mechanics can be found in \cite{preskill}.
In particular, a state is a {\em ray} in a Hilbert space, where a ray is
defined as an equivalence class of unit norm vectors that differ by 
multiplication by a nonzero complex scalar.
If we are looking at a subsystem of a larger quantum system, then the state
of the subsystem is not necessarily a ray.
If the state of the subsystem is a ray, then the state is called {\em pure}
and otherwise it is called {\em mixed}.
When we are considering these subsystems, the state of the system is 
represented by a {\em density operator}, i.e., a positive semi-definite
matrix with unit trace.
In the special case of a pure state, the density operator is the rank one
outer product of the corresponding ray with its conjugate transpose.
For a mixed state, the density operator is a convex combination of the
density operators of two or more pure states.

A discrete, memoryless quantum information source is an ensemble of
density operators $\rho_1 , \, \dots , \, \rho_M$ emitted with probabilities
$\alpha_1 , \, \dots , \, \alpha_M$.
Each density operator corresponds to a pure or a mixed state.
The goal of the quantum data compression problem formulated in 
\cite{schumacher} is to compress a sequence of pure quantum states into
the smallest possible Hilbert space with arbitrarily good reconstruction
fidelity in the limit as the sequence length approaches infinity.
In the special case where the ensemble consists of only pure states,
the problem has been solved in \cite{schumacher}, \cite{jozsa}, 
\cite{barnum96}.
The more general problem where the ensemble contains at least one mixed state
was first mentioned in \cite{jozsa94}.
In this case, the optimal compression rate is unknown \cite{horodecki}, 
\cite{horodecki99}, \cite{barnum}, but these papers provide upper and lower 
bounds on the best achievable compression rates.

When the matrices corresponding to the density operators for an ensemble of
mixed and/or pure states  commute, the quantum compression problem has been 
reformulated in \cite{barnum} as an equivalent classical information theory 
problem in which probability distributions are compressed and communicated.
Our analysis will be in terms of this formulation.
The problem of optimal mixed state coding has been considered in two different
scenarios.
In the first case, called the {\em visible source case}, 
the encoder knows the precise
sequence of states or probability distributions that it is transmitting.
In the second case, called the {\em hidden source case}, 
the encoder only has access to a measurement or ``side information'' sequence.
Each entry of this second sequence is found by taking a measurement of the
corresponding state; i.e., taking one experimental outcome of the probability 
distribution of the analogous entry in the original sequence.
Elsewhere in the quantum information literature this is called the 
{\em blind case}, but
the terminology ``hidden'' is more standard in the communications literature.
References \cite{horodecki}, \cite{horodecki99}, and \cite{barnum}
provide lower and upper bounds for the optimal rate of asymptotically
faithful compression which apply to both variants of the problem.

We provide a rate distortion interpretation of the problem which unifies the
analysis of both variants and leads
to the exact optimal rates for both the visible and blind versions.
Furthermore, the rate distortion framework leads to a natural generalization 
of the quantum compression problem in which the expected fidelity of 
reconstruction is asymptotically bounded from below but is not necessarily 
perfect.
To our knowledge, this problem has not been addressed earlier in the 
literature.
Our techniques provide the optimal compression rate for the both the 
visible and blind commuting cases in this setting.

It has come to our attention that \cite{ignacio} presents an alternate
proof of the achievability of the lower bound in the visible, commuting case 
where reconstruction is asymptotically perfect.

\subsection{Transmitting Probability Distributions}
\label{subsec:transmitPD}
Suppose that we have an ensemble of $M$ states with the corresponding 
discrete probability mass
functions $P_1, P_2, \dots , P_M$ that assume outcome values from the alphabet
${\cal Y} = \{1, \dots , N\}$.
We represent the alphabet $\{ 1, \dots, M \}$ by ${\cal X}$.
Let $p_{i,j}, \; i \in {\cal X}, \; j \in {\cal Y}$ denote the probability 
that a measurement of the $i^{\mbox{th}}$ state leads to outcome value $j$.
Hence, $p_{i,j} \; \geq \; 0, \; i \in {\cal X}, \; j \in {\cal Y},$
and $\sum_{j=1}^{N} p_{i,j} \; = \; 1, \; i \in  {\cal X} .$

Assume there is a memoryless source emitting a sequence of the mass functions.
In other words, there is a probability distribution on ${\cal X}$
and with probability $\alpha_i$ the source emits state $i$.
The source simultaneously produces a second sequence on ${\cal Y}$
which can be viewed as side information.
When the source emits state $i$, it also emits a side-information output
symbol $j \in {\cal Y}$ with probability $p_{i,j}$.
Let $\{X_\ell\}_{\ell \geq 1}$ and $\{Z_\ell\}_{\ell \geq 1}$ be the
output of the source corresponding to the sequence of states and the
sequence of side information, respectively.
For the original problem posed in \cite{barnum},
we wish to consider codes in which a receiver that knows the source model
generates a sequence $\{Y_\ell\}_{\ell \geq 1}$ of output values that fall in the 
``strongly typical set'' (see, e.g., \cite[\S 13.6]{ct})
for the state sequence $\{X_\ell\}_{\ell \geq 1}$.
More specifically, for each state $i$ the relative frequencies of 
the $N$ output symbols corresponding to the positions where $i$ is the
state emitted from the source should asymptotically converge to the 
probability mass function $P_i$ with probability $1$.
In other words, we measure the fidelity of the output sequence 
$\{Y_\ell\}_{\ell \geq 1}$ through the empirical distribution of sequences of pairs
$\{(X_\ell,Y_\ell)\}_{\ell \geq 1}$.
In practice, coding is performed from finite strings $X^L = X_1, X_2, \dots ,
X_L$ to output strings $Y^L = Y_1 , Y_2 , \dots , Y_L$.
Pick a block length $L$ and let $P^e_{X^LY^L}(i,j)$ denote the sample 
frequency of state and output pairs $(X_l,Y_l)=(i,j)$ over the range
$l \in \{1, \dots , L \}$.
Then for the compression problem with asymptotically perfect reconstruction
we require the Bhattacharyya-Wootters overlap \cite[p. 9]{barnum} of
the true probabilities $\alpha_i \, p_{i,j}$ and the empirical
frequencies of the state and output pairs to be arbitrarily close to 1
in the limit as $L$ approaches infinity. More precisely, we choose our
code to satisfy the constraint
\begin{equation}
\mbox{Pr} \left( \left( \sum_{i=1}^{M} \sum_{j=1}^{N} 
\sqrt{\alpha_i \, p_{i,j} \, P^e_{X^LY^L}(i,j)} \right)^2 
< 1 - \ep \right) \; < \; \delta \label{eq:bw}
\end{equation}
for arbitrarily small positive constants $\delta$ and $\ep$ whenever $L$
is sufficiently large.
The code may use probabilistic processes for the encoding and/or decoding.
The objective of the encoder is to compress the state sequence as much as
possible.

The source model for this problem superficially resembles the composite
source models discussed in \cite[\S 6.1]{berger}.
The key difference is the reversal of what is viewed as the side information
sequence and what is viewed as the primary source sequence.
For this reason, the analysis techniques developed for that source coding
problem do not appear to apply to this setting.

There are two obvious upper bounds to the minimum average number of bits
per symbol required in the encoding.
One of these bounds applies to both the visible and the blind versions of
the compression problem and the other applies only to the visible case.
For the visible problem, the encoder may simply transmit the sequence 
$\{X_\ell\}_{\ell \geq 1}$ and the decoder may use the appropriate probability 
mass function every time it receives a state to generate the output 
sequence.
With this algorithm, the expected number of bits per symbol used by the encoder
can come arbitrarily close to the entropy \cite{shannon} of the state alphabet:
\begin{displaymath}
- \sum_{i=1}^{M} \alpha_i \log_2 \alpha_i.
\end{displaymath}
Another possibility for either the blind or the visible case is for the 
encoder to transmit the sequence $\{Z_\ell\}_{\ell \geq 1}$ and the decoder to use 
the sequence without modifying it.
The entropy of this sequence is
\begin{displaymath}
- \left( \sum_{i=1}^{M} \sum_{j=1}^{N} \alpha_i p_{i,j} \right) \log_2 
\left( \sum_{i=1}^{M} \sum_{j=1}^{N} \alpha_i p_{i,j} \right) .
\end{displaymath}
It is easy to find situations where both of these procedures are suboptimal.
Consider the case where the $M$ probability mass functions are identical,
$\alpha_i = 1/M$ for all states $i$, and $p_{i,j} = 1/N$ for all
pairs of states $i$ and output symbols $j$.
In this case, transmitting the sequence $\{X_\ell\}_{\ell \geq 1}$ will require
$\log_2 M$ bits per symbol on average and transmitting the sequence
$\{Z_\ell\}_{\ell \geq 1}$ will require $\log_2 N$ bits per symbol on average.
Here the optimal coding procedure for both the visible and blind versions
of the problem would be to have the encoder transmit nothing and the decoder 
generate independent and equiprobable output symbols.
This coding procedure uses the ideal of zero bits per symbol.

It is possible to modify the entropy upper bound for some sources to avoid
the simple counterexample above.
Suppose that there are two or more output symbols $j$ which have a ``common
randomness;'' i.e., for which the $p_{i,j}$ are equal for all $i \in {\cal X}$.
Then an encoding strategy would be to introduce an erasure symbol, to
replace all occurrences of output symbols with common randomness in 
$\{Z_\ell\}_{\ell \geq 1}$ with the erasure symbol, and to encode the resulting
sequence to its entropy.
The decoder will not modify the ordinary symbols, and when it sees an erasure
symbol it will generate a symbol of ``common randomness'' with the appropriate
conditional probability.
In the case where $p_{i,j} > 0$ for all pairs 
$(i,j) \in {\cal X} \times {\cal Y}$, we will show that for the blind version 
of the problem with asymptotically perfect fidelity it is impossible to do 
better than this modified entropy bound.  
Some additional care needs to be provided in specifying the solution for
the blind version of the problem when there are pairs
$(i,j) \in {\cal X} \times {\cal Y}$ with $p_{i,j} = 0$, but the solution
is in the form of a mutual information.

\cite{horodecki99} and \cite{barnum} prove that a lower bound to the optimal 
compression ratio for both versions of the problem with asymptotically perfect
fidelity
is the mutual information between the state alphabet and the output alphabet
\begin{equation}
\sum_{i=1}^{M} \sum_{j=1}^{N} \alpha_i p_{i,j} \log_2 p_{i,j}
- \left( \sum_{i=1}^{M} \sum_{j=1}^{N} \alpha_i p_{i,j} \right) \log_2 
\left( \sum_{i=1}^{M} \sum_{j=1}^{N} \alpha_i p_{i,j} \right) ,
\label{eq:bound} 
\end{equation}
but leaves open the question whether this lower bound is attainable in either
the visible or the blind variants.
We will establish that it is achievable for the visible version of the problem.

Our analysis takes advantage of the tools of rate distortion theory.
The quantum information literation thus far has focused upon the
Bhattacharyya-Wootters overlap (see (\ref{eq:bw})) as a way to measure the 
closeness of two probability distributions.
This overlap is non-negative and is equal to one exactly when the two 
probability distributions are identical.
An equivalent and opposite way to measure the closeness of two probability
distributions is to discuss their ``distance'' or the distortion generated
by approximating one by the other.
In this setting, perfect fidelity corresponds to zero distortion.
The Bhattacharyya-Wootters overlap can be converted into such a distortion
function by being subtracted from one.
There are many other examples of interesting distortion functions that appear
in the probability and classical information literature.
The advantage of this interpretation is that rate distortion theory has
been studied extensively since \cite{shannon59}.
We will show that there is a very simple way to formulate and solve the
problem of compressing probability distributions in the rate distortion
setting.
It is also straightforward to generalize these results to the case where the
reconstruction fidelity is imperfect.

\section{Preliminaries}
We begin with several basic information-theoretic definitions.
Suppose we have two discrete and finite random variables $X$
and $Y$ whose joint probability distribution is $P_{XY}$.
The {\em entropy} of $X$ and {\em conditional entropy} of $X$
given $Y$ are defined as (see \cite[Ch. 2]{ct})
\begin{align}
   H(X)   & = \sum_{x\in\rm{supp}(P_X)} -P_X(x) \log(P_X(x)),
   \nonumber \\
   H(X|Y) & = \sum_{(x,y)\in\rm{supp}(P_{XY})}
              -P_{XY}(x,y) \log(P_{X|Y}(x|y)),
   \nonumber
\end{align}
where $\rm{supp}(P_X)$ is the support of $P_X$, i.e., the
set of $x$ such that $P_X(x)>0$. As done here, we will continue
to write random variables with upper-case letters and values
they take on with lower-case letters.
The {\em informational divergence} between $P_X$ and $P_Y$
is defined as
\begin{align}
   D(\, P_X \,\|\, P_Y \,) & = \sum_{x\in\rm{supp}(P_X)}
   P_X(x) \, \log \left( \frac{P_X(x)}{P_Y(x)} \right),
   \nonumber
\end{align}
and we write $D(P_X\|P_Y)=\infty$ when there is an $x$ in
$\rm{supp}(P_X)$ such that $P_Y(x)=0$. The informational
divergence is also  called the ``information for discrimination,''
the ``relative entropy'' and the ``Kullback-Leibler distance''
\cite[p. 20]{ck}, \cite[p. 18]{ct}.
The {\em mutual information} between $X$ and $Y$ is
defined as
\begin{align}
   I(X;Y) & = D(\, P_{XY} \,\|\, P_X P_Y \,) \nonumber \\
          & = H(X) - H(X|Y) \nonumber \\
          & = H(Y) - H(Y|X). \nonumber
\end{align}
A well-known property of these quantities is that they
are all non-negative \cite[Ch. 2]{ct}. Furthermore, $D(P_X\|P_Y)=0$
if and only if $P_X(x) = P_Y(x)$ for all $x$ in $\rm{supp}(P_X)$.
This implies that $I(X;Y)=0$ if and only if $X$ and
$Y$ are statistically independent.
Two other important properties involving convexity are given
as lemmas.
\begin{lemma}[{\cite[p. 30]{ct}}]
   $D(P_X \| P_Y)$ is convex in the pair
   $(P_X,P_Y)$, i.e., if $(P_{X_\ell},P_{Y_\ell})$,
   $\ell=1,2,\ldots,L$, are pairs of distributions,
   then for any nonnegative $\lambda_\ell$ which sum to one
   we have
   \begin{align}
      \sum_{\ell=1}^L \lambda_\ell \,
      D\left( P_{X_\ell} \,\|\, P_{Y_\ell}\right)
      \ge D\left( \sum_{\ell=1}^L \lambda_\ell P_{X_\ell}
          \, \left\| \, \sum_{\ell=1}^L \lambda_\ell P_{Y_\ell} \right.
          \right).
      \label{eq:Dconvexity1}
   \end{align}
   Equivalently, we can view $D(P_X\|P_Y)$ as a function of
   $P_{XY}$ and say that $D(P_X\|P_Y)$ is convex in $P_{XY}$.
   \label{lemma:Dconvexity}
\end{lemma}
\noindent
Let $J$ be a random variable taking on the value $\ell$ with
probability $\lambda_\ell$, $\ell=1,\ldots,L$.
We can write (\ref{eq:Dconvexity1}) as
\begin{align}
   \E_J \left[ D\left( P_{X_J} \,\|\, P_{Y_J}\right) \right]
   \ge D\left( \E_J \left[ P_{X_J} \right] \,\|\,
               \E_J \left[ P_{Y_J} \right] \right)
   \label{eq:Dconvexity}
\end{align}
where $E_J[\,\cdot\,]$ denotes expectation with respect to the
random variable $J$. We will sometimes drop the subscript $J$
and write $\E[\,\cdot\,]$ if it is clear with respect to which
random variable we are taking expectations.

\begin{lemma}[{\cite[p. 31]{ct}}]
   The mutual information $I(X;Y)$ is concave in $P_X$
   when $P_{Y|X}$ is fixed, and convex in $P_{Y|X}$ when
   $P_X$ is fixed. In other words, we have
   \begin{align}
      \E_J \left[ I\left( X_J \,;\, Y_J \right) \right]
      \le I\left( \E_J\left[X_J\right] \,;\, \E_J\left[Y_J\right] \right)
      \nonumber
   \end{align}
   when $P_{Y_J|X_J}$ is the same for all $J$, and
   \begin{align}
      \E_J \left[ I\left( X_J \,;\, Y_J \right) \right]
      \ge I\left( \E_J\left[X_J\right] \,;\, \E_J\left[Y_J\right] \right)
      \label{eq:Iconvexity}
   \end{align}
   when $P_{X_J}$ is is the same for all $J$.
   \label{lemma:Iconvexity}
\end{lemma}
\noindent

Our distortion measures will be defined in terms of the
{\em empirical probability distribution} of finite-length sequences
or strings.
The empirical probability distribution of the length-$L$ string
$x^L = x_1,x_2,\ldots,x_L$ with $x_\ell \in \cX$ is defined as
\begin{align}
   P^e_{x^L} (a) = \frac{n_{x^L}(a)}{L}
   \quad \mbox{for all } a \in \cX,
   \nonumber
\end{align}
where $n_{x^L}(a)$ is the number of occurrences of the letter
$a$ in the string $x^L$ \cite[p. 29]{ck},\cite[p. 279]{ct}.
A simple yet important property of $P^e_{x^L}$ is given by the
following lemma.
\begin{lemma}
   \begin{align}
      \E_{X^L} \left[ P^e_{X^L} \right] =
      \frac{1}{L} \sum_{\ell=1}^L P_{X_\ell}.
   \end{align}\
   \label{lemma:Empirical}
\end{lemma}
\begin{proof}
   We have, for all $a \in \cX$,
   \begin{align}
      \E_{X^L} \left[ P^e_{X^L}(a) \right] & =
         \E_{X^L} \left[ \frac{n_{X^L}(a)}{L} \right]
      \nonumber \\
      & = \frac{1}{L} \E_{X^L} \left[ \sum_{\ell=1}^L 1(X_\ell=a) \right]
      \nonumber
   \end{align}
   where $1(\,\cdot\,)$ is the indicator function that is 1 if its
   argument is true and is 0 otherwise. Since the expectation of
   a sum is the sum of the expectations~\cite[p. 10]{G96} , we have
   \begin{align}
      \frac{1}{L} \E_{X^L} \left[ \sum_{\ell=1}^L 1(X_\ell=a) \right]
      & = \frac{1}{L} \sum_{\ell=1}^L  
          \E_{X^L} \left[ 1(X_\ell=a) \right]
      \nonumber \\
      & = \frac{1}{L} \sum_{\ell=1}^L P_{X_\ell}(a).
      \nonumber
   \end{align}
\end{proof}

\subsection{Rate Distortion Theory}
\begin{figure}[t]
  \centerline{\includegraphics[scale=0.6]{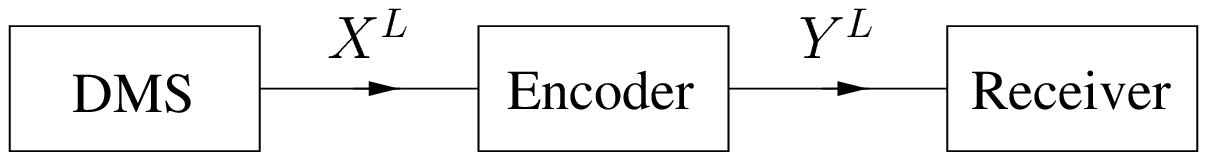}}
  \caption{Model for the rate distortion problem. }
  \label{fig:RDmodel}
\end{figure}

We describe the rate distortion problem as considered by
Shannon \cite{shannon59} (see Fig.~\ref{fig:RDmodel}). A discrete
memoryless source (DMS) produces a message string $X^L$ of $L$ 
independent and identically distributed letters from a
finite alphabet $\cX$. $X^L$ is encoded into one of $K=2^{LR}$
received strings $Y^L$ of $L$ letters from a finite alphabet
$\cY$. The rate of the encoder is thus $R$ bits per letter,
because one can represent any $y^L$ by a string of $LR$ bits.
There is a distortion measure $d(\cdot,\cdot)$ that associates
a non-negative number $d(x,y)$ with each pair $(x,y)$ of message
letter $x$ and receive letter $y$. The distortion between the
strings $x^L$ and $y^L$ is defined as the average of the
letter-to-letter distortions:
\begin{align}
   d(x^L,y^L) = \frac{1}{L} \sum_{\ell=1}^L d(x_{\ell},y_{\ell}),
   \nonumber
\end{align}
where we have abused notation by using the same symbol $d$ for
the letter-to-letter and string distortions.
Shannon generalized the letter-to-letter distortion measure in
\cite{shannon59}, but we will not be using that generalization here.

The fundamental problem of rate distortion theory is to
determine the minimum code rate $R$ such that the average
distortion between $X^L$ and $Y^L$ is upper bounded by some
number $\Delta$. The {\em rate distortion function} $R(\Delta)$
is thus defined as the greatest lower bound on $R$ such that
$\E[d(X^L,Y^L)] \le \Delta$. Shannon showed that $R(\Delta)$
has the simple form given by the following lemma.
\begin{lemma}[Shannon \cite{shannon59}]
   The rate distortion function of a DMS with distribution $P_X$
   and letter-to-letter distortion measure $d(\cdot,\cdot)$ is
   \begin{align}
      R(\Delta) & = \min\begin{Sb} P_{Y|X}: \\
                    \E[d(X,Y)] \le \Delta \end{Sb}
               I(X;Y). \nonumber
   \end{align}
\end{lemma}
\noindent
The achievability of the rate distortion function is usually
demonstrated by choosing a {\em random code} as follows:
the $L$ letters of each of the $2^{LR}$ code words are chosen
independently using $P_Y$. One then associates the ``typical''
strings $x^L$, i.e., those $x^L$ for which $P^e_{x^L}$ is close
to $P_X$, with one of the code words $y^L$ for which
$P^e_{x^Ly^L}$ is close to $P_{XY}$, where $P^e_{x^Ly^L}$ is
the empirical distribution of the $L$ pairs $(x_\ell,y_\ell)$.
One can show that if $R>R(\Delta)$ and $L$ is large, such a
code word $y^L$ exists and $d(x^L,y^L) \le \Delta$ with high
probability.

\begin{figure}[t]
  \centerline{\includegraphics[scale=0.6]{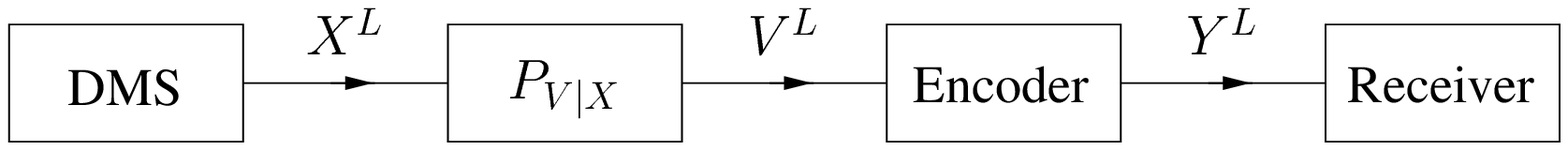}}
  \caption{Model for the rate distortion problem with a hidden source.}
  \label{fig:RDmodel_hidden}
\end{figure}

A generalization of the rate distortion problem was given
by Dobrushin and Tsybakov in \cite{dt}
(see Fig.~\ref{fig:RDmodel_hidden}).
The new twist is that the encoder sees only a noisy version
$V^L$ of the message $X^L$, where $v_\ell$ is generated by
$x_\ell$ via the memoryless channel $P_{V|X}(v_\ell|x_\ell)$
for all $\ell$. The DMS is sometimes called a
``remote source'' \cite[p. 78]{berger},\cite[p. 136]{ck}.
We will call the DMS a {\em hidden source}, $X^L$ the
{\em hidden source string}, $V^L$ the
{\em visible source string} and $P_V$ the
{\em visible distribution}. Note that if $V=X$ we have
the original rate distortion problem.

The goal is again to determine the minimum code rate $R$ such
that the average distortion between $X^L$ and $Y^L$ is upper
bounded by some number $\Delta$. The rate distortion function
$R(\Delta)$ is thus defined as before, and
Dobrushin and Tsybakov proved the following lemma.
\begin{lemma}[Dobrushin and Tsybakov \cite{dt}]
   The rate distortion function of a hidden DMS with
   distribution $P_X$, visible distribution $P_V$, and
   single-letter distortion measure $d(\cdot,\cdot)$ is
   \begin{align}
      R(\Delta) & = \min\begin{Sb} P_{Y|V}: \\
                   \E[d(X,Y)] \le \Delta \end{Sb}
               I(V;Y). \nonumber
   \end{align}
\end{lemma}
\noindent
Note that
\begin{align}
   I(V;Y) & =    H(Y) - H(Y|V)  \nonumber \\
          & =    H(Y) - H(Y|VX) \nonumber \\
          & \ge  H(Y) - H(Y|X)  \nonumber \\
          & = I(X;Y), \nonumber
\end{align}
where the second equality follows because $Y$ is independent
of $X$ given $V$, and the inequality follows because
conditioning cannot increase entropy \cite[p. 27]{ct}.
Thus, not surprisingly, the best rate when $X^L$ is
hidden is at least as large as when $X^L$ is visible.

\begin{lemma}
   The random variables of the rate distortion problem with
   a hidden source satisfy
   \begin{align}
      H(Y^L) \ge I(V^L;Y^L)
             \ge \sum_{\ell=1}^L I(V_{\ell};Y_{\ell})
             \ge L \cdot I\left(\Vbar;\Ybar\right),
      \label{eq:Ibounds}
   \end{align}
   where $P_{\Vbar \, \Ybar} = \sum_{\ell=1}^L P_{V_\ell Y_\ell}/L$.
   \label{lemma:Ibounds}
\end{lemma}
\begin{proof}
   The first inequality follows by the non-negativity of $H(Y^L|V^L)$.
   In fact, $Y^L$ is usually a function of $V^L$ so that
   $H(Y^L|V^L)=0$ and $H(Y^L)=I(V^L;Y^L)$.
   The second inequality follows by
   \begin{align}
      I(V^L;Y^L) & = \sum_{\ell=1}^L H(V_\ell|V^{\ell-1})
                     - H(V_\ell | Y^L V^{\ell-1})
      \nonumber \\
                 & = \sum_{\ell=1}^L H(V_\ell) - H(V_\ell | Y^L V^{\ell-1})
      \nonumber \\
                 & \ge \sum_{\ell=1}^L H(V_\ell) - H(V_\ell | Y_\ell).
      \nonumber
   \end{align}
   The third inequality follows by viewing the sum over
   the $I(V_{\ell};Y_{\ell})$ as $L$ times $\E_J[I(V_J;Y_J)]$,
   where $J$ takes on the value $\ell$ with probability $1/L$ for
   $\ell=1,\ldots,L$. The bound (\ref{eq:Iconvexity}) then gives
   the desired result.
\end{proof}

\section{Quantum Rate Distortion}
\begin{figure}[t]
  \centerline{\includegraphics[scale=0.6]{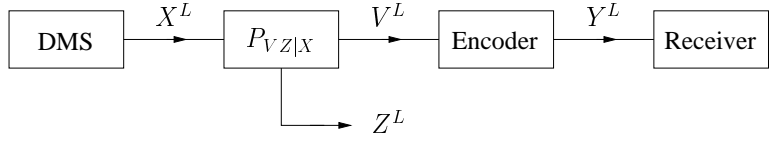}}
  \caption{Model for the quantum rate distortion problem.}
  \label{fig:Qmodel}
\end{figure}

We deal with the visible and hidden (or blind) source problems
simultaneously by introducing an auxiliary string $Z^L$
to the model of Fig. \ref{fig:RDmodel_hidden}
(see Fig. \ref{fig:Qmodel}). $Z^L$ represents the outcomes
of measurements and is called side information in
Section \ref{subsec:transmitPD}.
The terms of $Z^L$ take on values in the alphabet $\cZ$ and
are generated together with $V^L$ as outputs of a memoryless
channel $P_{VZ|X}$.

We are interested in string distortion measures
$d(\cdot,\cdot)$ that depend on $(x^L,y^L)$ only through
the empirical distribution $P^e_{x^Ly^L}$. Thus, with some
abuse of notation we can write $d(x^L,y^L)=d(P^e_{x^Ly^L})$.
For example, using (\ref{eq:bw}) the Bhattacharyya-Wootters
distortion measure could be defined as
\begin{align}
   d(P^e_{x^Ly^L}) & = 1 - \left[
      \sum_{x\in\cX,z\in\cZ}
      \sqrt{ P_{XZ}(x,z) \, P^e_{x^Ly^L}(x,z) }
   \right]^2,
   \label{eq:bwd}
\end{align}
where $Z$ plays the role of the measurement outcomes in
Section \ref{subsec:transmitPD}. The visible case has $V=X$
while the hidden case can have $V\ne X$ and has $V=Z$.
As a second example, a natural information-theoretic
distortion measure is the informational divergence
\begin{align}
   d(P^e_{x^Ly^L}) & = D\left(
      P^e_{x^LZ} \| P^e_{x^Ly^L} \right),
   \label{eq:idd}
\end{align}
where $P^e_{x^LZ}(a,b)$ is defined as
$\left[\sum_{c\in\cY} P^e_{x^Ly^L}(a,c)\right] P_{Z|X}(b|a)$
for all $a\in\cX$ and $b\in\cZ$, i.e.,
$P^e_{x^LZ} = P^e_{x^L}\, P_{Z|X}$. Observe that low distortion is
achieved only if the empirical distribution of $(x^L,y^L)$
is close to the desired distribution $P^e_{x^LZ}$.

We next impose an additional restriction on $d(\cdot)$, namely
that $d(P_{XY})$ be convex in $P_{XY}$, i.e.,
\begin{align}
   \E_J \left[ d\left( P_{X_J Y_J} \right) \right]
   \ge d\left( \E_J \left[ P_{X_J Y_J} \right] \right),
   \label{eq:dconvexity}
\end{align}
where $J$ is a finite random variable. The distortion measure
(\ref{eq:idd}) meets this requirement by Lemma \ref{lemma:Dconvexity}.
The distortion measure (\ref{eq:bwd}) also meets this requirement
since, for $\lambda_\ell\ge 0$ and $\sum_\ell \lambda_\ell=1$,
we have
\begin{align}
   1-\left[ \sum_{x,z} \sqrt{ \sum_{\ell}
      \lambda_\ell a_\ell(x,z) } \right]^2
   & \le
   1- \sum_{\ell} \left[ \sum_{x,z} \sqrt{
      \lambda_\ell a_\ell(x,z) } \right]^2 \nonumber \\
   & = \sum_{\ell} \lambda_\ell \left\{ 1 -
       \left[ \sum_{x,z} \sqrt{a_\ell(x,z)} \right]^2 \right\},
   \nonumber
\end{align}
where $a_\ell(x,z)=P_{XZ}(x,z) \, P^e_{X_\ell Y_\ell}(x,z)$ and
where the first step follows by Minkowski's inequality \cite[p. 523]{G}.

We call the problem of finding the rate distortion function
for this set-up the {\em quantum commuting density operator}
(quantum CDO) rate distortion problem.
The following lemma gives a lower bound on the rate distortion
function.
\begin{lemma}[Rate Lower Bound]
   The rate $R$ of the quantum CDO rate distortion problem
   with expected distortion
   $\E \left[d\left(P^e_{X^LY^L}\right)\right] = \Delta$
   satisfies
   \begin{align}
      R & \ge \min\begin{Sb} P_{Y|V}: \\
                  d(P_{XY} ) \le \Delta \end{Sb}
              I(V;Y).
      \label{eq:lowerBound}
   \end{align}
   \label{lemma:lowerBound}
\end{lemma}
\begin{proof}
   A simple upper bound on $H(Y^L)$ is the logarithm of the
   number of possible values $Y^L$ takes on with nonzero
   probability \cite[Sec. 6]{shannon}, i.e., the logarithm of
   the number of code words. We thus have
   \begin{align}
      R \ge H(Y^L)/L
        \ge I(\Vbar;\Ybar)
        \ge \min\begin{Sb} P_{Y^L|V^L}: \\
             \E[d(X^L,Y^L)] \le \Delta \end{Sb} I(\Vbar;\Ybar),
      \nonumber
   \end{align}
   where the second inequality follows by (\ref{eq:Ibounds}),
   and the third inequality because of the minimization.
   Next, we have
   \begin{align}
      \E\left[d(P^e_{X^LY^L})\right] &
      \ge d\left( \E\left[ P^e_{X^LY^L} \right] \right)
      = d\left( P_{\Xbar\,\Ybar} \right),
      \nonumber
   \end{align}
   where $P_{\Xbar\,\Ybar} = \sum_{\ell=1}^L P_{X_\ell Y_\ell}/L$.
   The inequality follows by the convexity of $d(\cdot)$ and
   the equality by Lemma~\ref{lemma:Empirical}.
   Thus, the condition $\E[d(X^L,Y^L)] \le \Delta$
   implies that $d(P_{\Xbar\,\Ybar}) \le \Delta$, and we have
   \begin{align}
      R \ge  \min\begin{Sb} P_{Y^L|V^L}: \\
             d(P_{\Xbar\,\Ybar}) \le \Delta \end{Sb} I(\Vbar;\Ybar).
      \nonumber
   \end{align}
   This is the same as (\ref{eq:lowerBound}) because the
   minimization over $P_{Y^L|V^L}$ is the same as the minimization
   over $P_{\Ybar\left|\Vbar\right.}$.
\end{proof}

We next show that the lower bound of
Lemma~\ref{lemma:lowerBound} can be approached arbitrarily
closely, and is thus the desired rate distortion function.
\begin{lemma}[Achievable Rates]
   For any $\delta>0$ and distortion $\Delta$ there exists a
   block code of sufficiently large block length for which
   \begin{align}
      R & \le \min\begin{Sb} P_{Y|V}: \\
                  d(P_{XY}) \le \Delta \end{Sb}
              I(V;Y) + \delta. \nonumber
   \end{align}
   \label{lemma:achievableRates}
\end{lemma}
\begin{proof}
   We give only a very brief sketch of the proof for this
   preliminary version of the paper. \\
   The code is generated by choosing some $P_{Y|V}$ and then
   randomly choosing each symbol of the $2^{LR}$ code words
   independently using the resulting $P_Y$.
   Let the $k$th code word be $y_k^L$ and choose some $\epsilon>0$.
   For each $v^L$ satisfying $|P^e_{v^L}(a)-P_V(a)|\le\epsilon$
   for all $a$, one looks for a code word $y_k^L$ such that
   $|P^e_{v^Ly_k^L}(a,b)-P_{VY}(a,b)|\le\epsilon$ for all $a$ and $b$.
   Let $\cE_k(v^L)$ be the event that the $k$th code word $Y_k^L$,
   now regarded as a random variable, is such a code word.
   Lemma 13.6.2 in \cite[p. 359]{ct} assures us that
   \begin{align}
      2^{-L\, [I(V;Y)+\epsilon_1]} \le \Pr\left[\cE_k(v^L)\right] \le
      2^{-L\, [I(V;Y)-\epsilon_1]}, \nonumber
   \end{align}
   where $\epsilon_1 \rightarrow 0$ as $\epsilon \rightarrow 0$
   and $L \rightarrow \infty$.
   Continuing as in \cite[Sec. 13.6]{ct}, one will need
   $K \approx 2^{L\,I(V;Y)}$ code words to ensure that
   $\cE_k(v^L)$ occurs for at least one $k$ for all the ``typical''
   $v^L$. One can also use the approach in \cite[Sec. 13.6]{ct}
   to show that the distortion criterion is met for each such
   $(v^L,y_k^L)$ pair with high probability.

   The code construction we have just described can be done for
   any $P_{Y|V}$, so we choose that $P_{Y|V}$ which minimizes
   the rate $I(V;Y)$.
\end{proof}

\begin{theorem}
   The rate distortion function of the quantum CDO rate distortion
   problem is
   \begin{align}
      R(\Delta) & = \min\begin{Sb} P_{Y|V}: \\
                   d(P_{XY}) \le \Delta \end{Sb}
               I(V;Y). \nonumber
   \end{align}
   \label{theorem:quantumRD}
\end{theorem}

\subsection{Examples}

We give examples to demonstrate how one can apply the
above results. Consider Example 8 of \cite{barnum}
in which the states $\rho_1 = \diag(\alpha_1,1-\alpha_1)$
and $\rho_2 = \diag(\alpha_2,1-\alpha_2)$ have prior probabilities
$p$ and $1-p$, respectively, where $\diag(a,b)$ is
a diagonal matrix with entries $a$ and $b$. In
\cite{barnum} it is shown that one may regard the two states
as biased coins $c_1$ and $c_2$ that take on the values H
(for heads) with respective probabilities $\alpha_1$ and $\alpha_2$,
and the value T (for tails) with respective probabilities
$1-\alpha_1$ and $1-\alpha_2$. Adapting this to Fig. \ref{fig:Qmodel},
we let $X^L$ be the sequence of coins and $Z^L$ a sequence of
outcomes of coin tosses, i.e., $P_X(c_1)=p$, $P_X(c_2)=1-p$,
$P_{Z|X}(H|c_1)=\alpha_1$, $P_{Z|X}(H|c_2)=\alpha_2$, and so on.

\begin{figure}[t]
  \centerline{\includegraphics[scale=0.6]{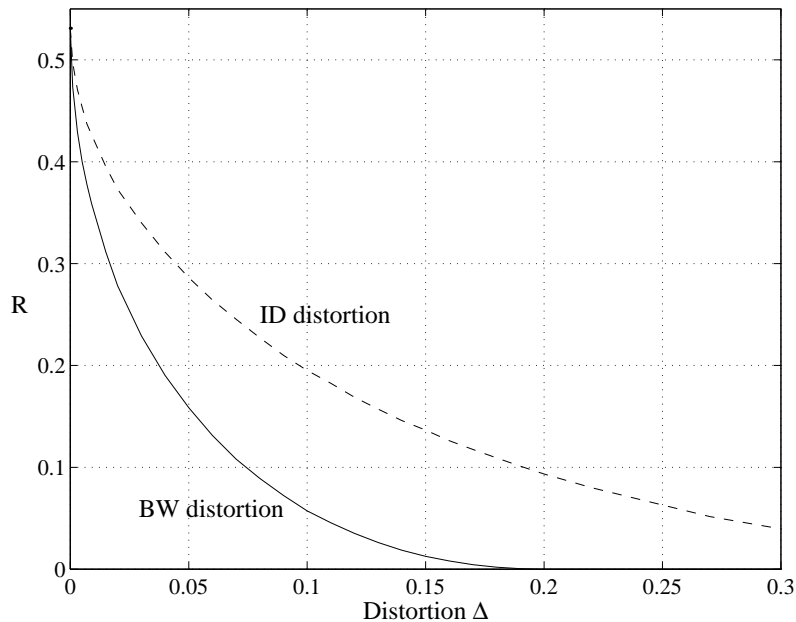}}
  \caption{Rate distortion function for a visible source.}
  \label{fig:visibleExample}
\end{figure}

Consider the visible case where $V=X$, so that the rate distortion
function is
\begin{align}
   R(\Delta) = \min\begin{Sb} P_{Y|X} : \\
               d(P_{XY} ) \le \Delta \end{Sb}
               I(X;Y). \nonumber
\end{align}
If $\Delta=0$ then $P_{XY}=P_{XZ}$ for both the
Bhattacharyya-Wootters and the informational divergence
distortion measures. Thus, we have
\begin{align}
   I(X;Y) = I(X;Z) =
   h\left(p\,\alpha_1+ (1-p)\,\alpha_2 \right) -
   \left[p\,h(\alpha_1) + (1-p)\,h(\alpha_2) \right],
   \nonumber
\end{align}
where $h(\alpha)=-\alpha\log_2(\alpha)-(1-\alpha)\log_2(1-\alpha)$
is the {\em binary entropy function} \cite[Fig. 7]{ct}.
For a concrete example, set $p=1/2$, $\alpha_1=1/10$ and
$\alpha_2=9/10$. Then $I(X;Y) \approx 0.5310$ is the ultimate limit on
data compression with no distortion; Fig. \ref{fig:visibleExample}
shows $R(\Delta)$ as a function of $\Delta$ for both the
Bhattacharyya-Wootters (BW) and informational divergence (ID)
distortion measures. Observe that $R(\Delta)$ is convex \cite{shannon59}.

\begin{figure}[t]
  \centerline{\includegraphics[scale=0.6]{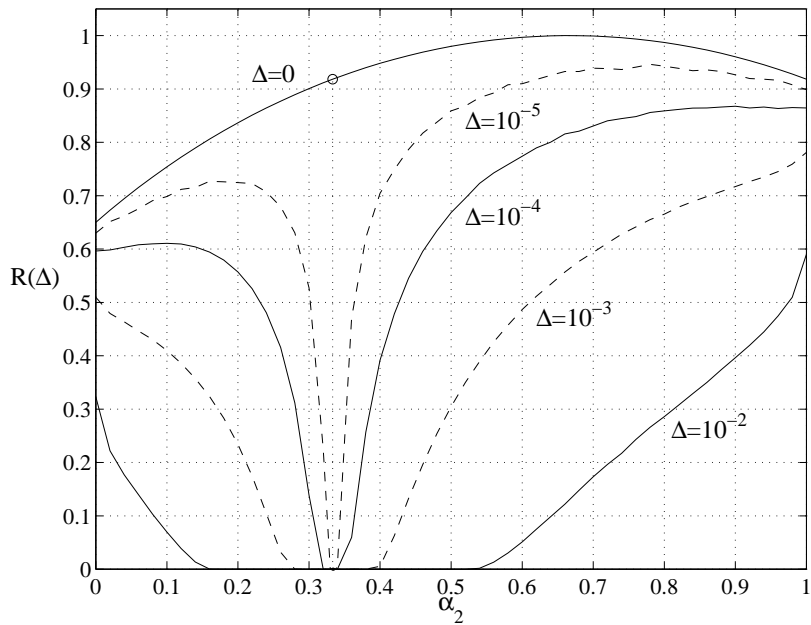}}
  \caption{Rate distortion function for a hidden source.}
  \label{fig:hiddenExample}
\end{figure}

Consider next the hidden source case (or blind case) where $V=Z$.
We thus have
\begin{align}
   R(\Delta) = \min\begin{Sb} P_{Y|Z} : \\
               d(P_{XY} ) \le \Delta \end{Sb}
               I(Z;Y). \nonumber
\end{align}
Again, if $\Delta=0$ then $P_{XY}=P_{XZ}$ for both the
Bhattacharyya-Wootters and the informational divergence
distortion measures. Performing the optimization,
we find that $R(0)$ can be a {\em discontinuous} function
of $\alpha_2$; for $\alpha_2 \ne \alpha_1$ we have
$R(0)=H(Z)=h(p\,\alpha_1+ (1-p)\,\alpha_2)$ and for
$\alpha_2=\alpha_1$ we have $R(0)=0$.
For example, suppose that $p=1/2$ and $\alpha_1=1/3$.
We plot $R(\Delta)$ as a function of $\alpha_2$ for various
$\Delta$ and the Bhattacharyya-Wootters distortion measure
in Fig. \ref{fig:hiddenExample}. Observe that as
$\Delta \rightarrow 0$ we will have a discontinuity at
$\alpha_2=1/3$. In practice, this discontinuity does
not occur because $\Delta=0$ is impossible for finite
block lengths. Furthermore, if $\Delta$ is not too
small, say $\Delta = 10^{-3}$, then for many
$\alpha_2$ one can achieve substantially better compression
rates than $R(0)$.

\section{Conclusions}
The problem of determining optimal compression limits for quantum information
has recently generated considerable interest.
In the special case of an ensemble of mixed states with commuting density
operators, we use rate distortion theory to find the optimal rates in 
both the visible and blind versions of the problem.
We also generalize this special case of the quantum compression problem
to the setting where the reconstruction is not faithful.

\section*{Acknowledgment}
We would like to thank C. Fuchs for bringing this problem to our
attention and V. Goyal for feedback on the manuscript.
We would also like to thank I. Cirac for providing us with a draft
of \cite{ignacio}.



\begin{thebibliography}{99}
\bibitem{shannon}
C. E. Shannon, ``A mathematical theory of 
communication,'' {\em Bell System Tech. J.} 27, 379-423, 623-656, 1948.

\bibitem{bennett}
C. H. Bennett and P. W. Shor, ``Quantum information theory,''
{\em I.E.E.E. Trans. Inform. Theory} 44, 2724-2742, 1998.

\bibitem{shannon59}
C. E. Shannon, ``Coding theorems for a discrete source with a fidelity
criterion,'' {\em I.R.E. National Convention Record} Part 4, 142-163, 1959.

\bibitem{preskill}
J. Preskill, http://www.theory.caltech.edu/people/preskill/ph229/\#lecture.

\bibitem{schumacher}
B. W. Schumacher, ``Quantum coding,'' {\em Phys. Rev. A} 51, 2738-2747, 1995. 

\bibitem{jozsa}
R. Jozsa and B. W. Schumacher, ``A new proof of the quantum noiseless coding
theorem,'' {\em J. Modern Optics} 41, 2343-2349, 1994.

\bibitem{barnum96} 
H. Barnum, C. A. Fuchs, R. Jozsa, and B. Schumacher,
``General fidelity limits for quantum channels,'' {\em Phys. Rev. A} 54, 
4707, 1996. 

\bibitem{jozsa94}
R. Jozsa, ``Quantum noiseless coding of mixed states,'' Talk given at
the Third Santa Fe workshop on Complexity, Entropy, and the Physics of
Information, May 1994.

\bibitem{horodecki}
M. Horodecki, ``Limits for compression of quantum information carried by
ensembles of mixed states,'' {\em Phys. Rev. A} 57, 3364-3369, 1998.

\bibitem{horodecki99}
M. Horodecki, ``Towards optimal compression for mixed signal states,'' 
LANL ArXiV.org e-print quant-ph/9905058, 1999. 

\bibitem{barnum} 
H. Barnum, C. M. Caves, C. A. Fuchs, R. Jozsa, and B. Schumacher,
``On quantum coding for ensembles of mixed states,'' 
LANL ArXiV.org e-print quant-ph/0008024, August 2000.

\bibitem{ignacio}
W. D\"ur, G. Vidal, and J. I. Cirac, ``Visible compression of commuting
density operators,'' to appear in LANL ArXiV.org.

\bibitem{ct}
T. M. Cover and J. A. Thomas, {\em Elements of Information Theory},
Wiley, New York 1991.

\bibitem{berger}
T. Berger, {\em Rate Distortion Theory: A Mathematical Basis for Data 
Compression}, Prentice-Hall, New Jersey 1971.

\bibitem{dt}
R. L. Dobrushin and B. S. Tsybakov, ``Information transmission with additional 
noise,'' {\em I.E.E.E. Trans. Inform. Theory} 8, 293-304, 1962.

\bibitem{ck}
I. Csisz\'ar and J. K\"orner, {\em Information Theory, Coding Theorems
for Discrete Memoryless Systems}, Akad\'emiai Kiad\'o, Budapest 1981.

\bibitem{G96}
R.G. Gallager, {\em Discrete Stochastic Processes},
Kluwer, Boston 1996.

\bibitem{G}
R. G. Gallager, {\em Information Theory and Reliable Communication},
Wiley, New York 1968.


\end{thebibliography}
\end{document}